%% file: arxiv-sub.tex
\documentclass[pre,twocolumn,floatfix]{revtex4}%

\input{macros.tex}

\usepackage{epsfig,amssymb,amsmath}
\usepackage{pstricks}

\newrgbcolor{lblue}{0.75 0.75 1.0}

\begin{document}

\title{Functionality Encoded In Topology? \\
Discovering Macroscopic Regulatory Modules From \\
Large-Scale Protein-DNA Interaction Networks}

\author{Riccardo Boscolo}
\email{riccardo@ee.ucla.edu}
\affiliation{Department of Electrical Engineering \\ University of California, Los Angeles}
\author{Behnam A. Rezaei}
\email{behnam@ee.ucla.edu}
\affiliation{Department of Electrical Engineering \\ University of California, Los Angeles}
\author{P. Oscar Boykin}
\email{boykin@ece.ufl.edu}
\affiliation{Department of Electrical and Computer Engineering \\ University of Florida, Gainesville}
\author{Vwani P. Roychowdhury}
\email{vwani@ee.ucla.edu}
\affiliation{Department of Electrical Engineering \\ University of California, Los Angeles}


\begin{abstract}

The promise of discovering a functional blueprint of a cellular system from large-scale and high-throughput sequence and experimental data is predicated on the belief that the same \emph{top-down} investigative approach that proved successful in other biological problems (e.g. DNA sequencing) will be as effective when it comes to inferring more complex intracellular processes. The results in this paper address this fundamental issue in the specific context of transcription regulatory networks. In particular, we consider a recently introduced experimental technique, the genome-wide location analysis for DNA-binding regulators, which allows the construction of network topologies relating transcriptional regulators with all DNA promoter regions they are capable of interacting with. Although simple recurring regulatory motifs have been identified in the past, due to the size and complexity of the connectivity structure, the subdivision of such networks into larger, and possibly inter-connected, regulatory modules is still under investigation. Specifically, it is unclear whether functionally well-characterized transcriptional sub-networks can be identified by solely analyzing the connectivity structure of the overall network topology. In this paper, we show that transcriptional regulatory networks can be systematically partitioned into communities whose members are consistently functionally related. We applied the partitioning method to the transcriptional regulatory networks of the yeast \emph{Saccharomyces cerevisiae}; the resulting communities of gene and transcriptional regulators can be associated to distinct functional units, such as amino acid metabolism, cell cycle regulation, protein biosynthesis and localization, DNA replication and maintenance, lipid catabolism, stress response and so on. Moreover, the observation of inter-community connectivity patterns provides a valuable tool for elucidating the inter-dependency between the discovered regulatory modules.
\end{abstract}

\maketitle

\section{Introduction}
\subsection{Motivation and Background: Bottom-Up Vs. Top-Down}
We address one of the primary goals of systems biology: Discovering
a functional blueprint of a cellular system from large-scale and high-throughput sequence and experimental data. Such a blueprint would describe how the different components (e.g., genes, proteins, signaling molecules etc.) work together to perform various tasks in the cell. A  wealth of information, obtained through decades of ingenious but painstaking   investigations by biochemists and biologists, have helped elucidate many aspects and components of the complex functional organization in different types of cells and organisms. This investigative approach can be broadly described as a {\em bottom-up approach}, where several smaller components and systems are first modeled under carefully constrained conditions; larger systems with increasing complexity, and comprising well-studied smaller components, are then characterized in subsequent steps.

In a sharp contrast to this established approach, recently introduced high-throughput experimental techniques hold the promise of enabling a {\em top-down} investigation. The whole-genome shotgun sequencing method\cite{venter-nature}\cite{venter-science}, where thousands of short strands of DNA are sequenced in parallel and then pieced together in a post-processing computational step to reconstruct a complete genome sequence, provides a good example of the few successes that have fueled high expectations. In order to extend this trend to functional investigations of cellular systems, new experimental and analysis tools are being designed. Typically, the simultaneous average activity or interaction levels  of thousands of indicator molecules and agents (e.g., genes, proteins, and signaling molecules) are observed or tracked in a population of  cells that have been subjected to different conditions, or have been otherwise manipulated with. Whole-genome DNA microarray assays, which can estimate the expression levels of thousands of genes, constitute a prime example of such a technology.  These observed profiles are then processed, and statistically significant dependencies, correlations, and other structural relationships inherent in the data sets are determined. These inherent dependencies among the observed agents are then expected to yield working hypotheses about functional relationships among them; the resulting hypotheses can then be investigated further using tailored experiments to obtain a more detailed description of the functional blocks and mechanisms.

This top-down approach, however, has yet to prove its usefulness when it comes to inferring intracellular mechanisms and processes. A number of studies have attempted to combine  sequence information and experimental data and have devised methods for determining potential functional blocks or hidden regulatory mechanisms \cite{ihmels-nature,segal-NG,hughes-rosetta,lee-science,pnas-nca,gardner-science}. For example, partitioning of genes into clusters (based on similarity of their activities or profiles and related sequence information) could lead one to hypothesize that genes in the same cluster are part of the same pathway, or that their profiles constitute a signature of a particular functionality in the cell. Such advances notwithstanding, {\em basic questions} concerning the power and usefulness of these large-scale approaches {\em are yet to be resolved}. {\em First}, given the complexity of cellular processes, the number of agents/outputs tracked in any large-scale study comprises only a small fraction of all the participating factors or agents. {\em Second}, the final data sets  are the outputs of a highly  complex regulatory process involving interactions among a large number of bio-molecules that operate at different  stages and different parts of the cell. Clearly, such considerations lead to several basic questions: {\em Is there enough information} in these data sets to be able {\em to formulate sufficiently many significant hypotheses}, which  would ultimately lead to a detailed reconstruction of functional blocks?  If not, then {\em how much prior knowledge} would one need to incorporate before one has {\em enough information}?

\subsection{Approach and a Preview of Results}
The results in this paper {\em address the above-mentioned fundamental issues} in the {\em specific context of transcription regulatory networks}. In particular, we consider a recently introduced experimental technique, the {\em genome-wide location analysis} of DNA-binding regulators\cite{lee-science}, which allows one to construct an interaction network between regulatory proteins (also referred to as transcription regulators or factors) and genes. The experiment relates any given transcriptional regulator with all DNA promoter regions they are capable of interacting with. Typically, the resulting networks involve several thousands of nodes (i.e., hundreds of regulators and thousands of genes), and an even larger number of edges, each representing a physical interaction; that is, a node representing a regulator is connected to a node representing a gene by an edge, if the corresponding protein binds to the promoter region of the corresponding gene with a high confidence level\footnote{The authors of~\cite{harbison} describe in their paper a method for augmenting the network structure to include a larger set of interactions, with the aid of expression data analysis. However, in this work we restricted our analysis to the topology directly derived from the immunoprecipitation experimental data.}. While the design of these large-scale experiments need {\em genome sequence information} (e.g., prior knowledge about candidate regulatory proteins, and the location of genes and their corresponding promoter regions in the genome), {\em no prior knowledge about any functionality of the genes or the regulators} is needed.  The functionality-blind design of these experiments makes the inferred transcriptional regulatory networks good candidates for  answering some of the basic questions raised in the preceding paragraph: \\ \\
Can {\em functionally well characterized transcriptional sub-networks} be identified by {\em solely analyzing the connectivity structure} of the overall network topology? In particular, we aim at establishing whether partial or complete cell pathways can be {\em automatically recognized} by a method that {\em relies exclusively on the interaction network}, with {\em no other prior knowledge} about the specific organism under study.
\\ \\
In our analysis, we applied the Girvan and Newman (GN) community partitioning algorithm~\cite{newman-girvan} to the transcriptional regulatory networks of the yeast \emph{Saccharomyces cerevisiae} (\emph{S.cerevisiae}) (details on the data are provided in the Results section). This partitioning method relies purely on the amount of information held within the connectivity structure of the transcriptional network, and does not require the introduction of specific parameters describing the modules sought after. The GN algorithm returns a nested set of "communities" or "groups" of genes in the form of a tree structure, where a {\em community} is characterized by the fact that nodes within the community  are densely inter-connected, while connections are significantly sparser between members of different communities. For example, the transcription network of \emph{S.cerevisiae} is divided into {\em 15 different communities at the first level}. These communities are then subdivided into further communities {\em in a nested fashion}. The depth of the resulting tree decomposition and the total number of communities are determined by a parameter called the "modularity index", and its definition and issues related to its choice are discussed in more detail in later sections.

In a top-down approach, each such community would comprise a set of hypotheses, which would then need to be tested using further experiments involving the genes belonging to the same community, or exploiting homology with related organisms. This approach, however, can  {\em succeed} only if these {\em structure-based communities} have {\em coherent functional themes}.  Since {\em S.cerevisiae} is a  fairly well studied organism, we can obviate the need for experimentation and verify whether the communities represent functional blocks based on the information in existing literature. Toward this end, we performed {\em an automated search of the Gene Ontology (GO) database}\cite{go-ref}, and obtained  a functional annotation of the communities, along with their {\em significance levels.}

\begin{figure*}[t]
\centerline{\includegraphics[width=6in]{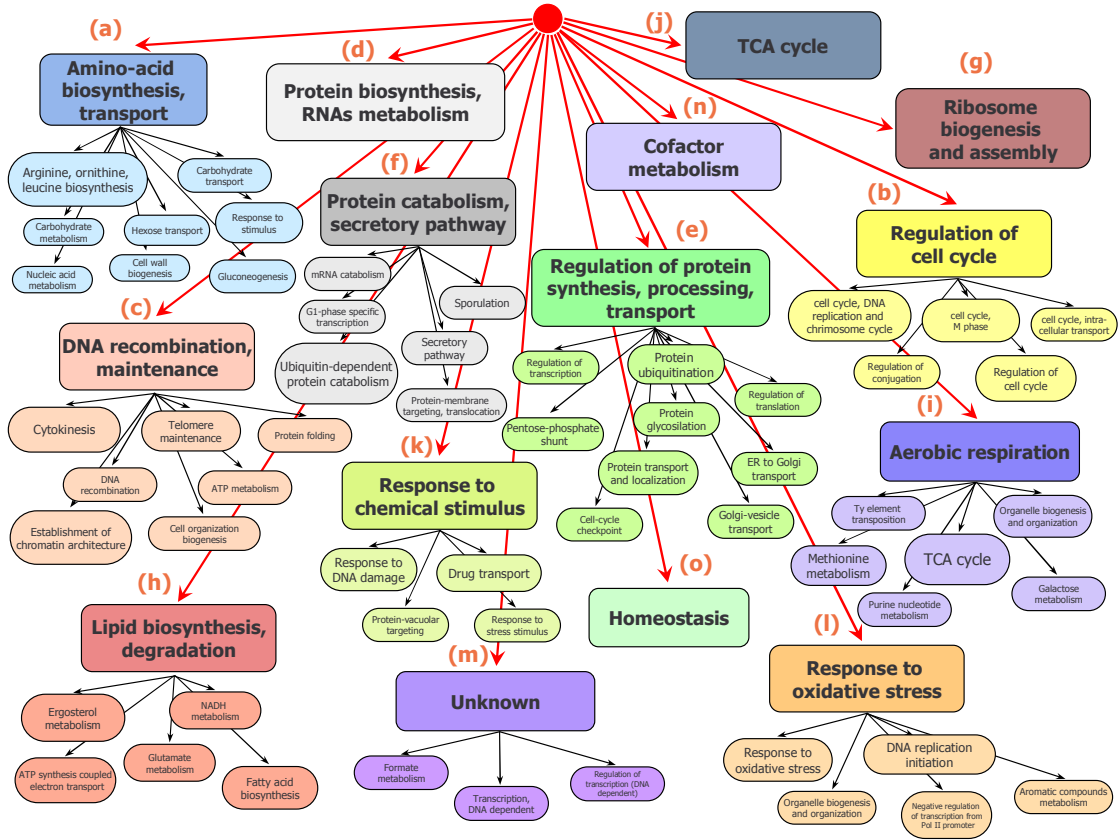}}
\caption{Tree structure organization of the functional modules obtained by partitioning the transcriptional regulatory network of \emph{S.cerevisiae} (only the top two levels of partitioning are shown). Details on the characteristics of each regulatory sub-network are provided in the Results section.}
\label{com-tree-struct}
\end{figure*}

When the communities  are tagged with their corresponding statistically-significant functional annotations,  it results in {\em a remarkable functional blue-print} of the cell (See Fig.~\ref{com-tree-struct}): At the bottom level of the nested organizational architecture,  genes and regulators are grouped into homogeneous modules performing basic functionalities; these basic functional modules are in turn organized {\em into a hierarchy} that capture  progressively higher levels of functionality culminating into high-level {\em cellular operational blocks}, such as  {\em amino acid metabolism}, {\em cell cycle regulation}, {\em protein biosynthesis} and localization, {\em DNA replication} and maintenance, {\em lipid catabolism}, {\em stress response} and so on. In addition, {\em the patterns of inter-community edges}  provide important insights into how large {\em regulatory modules are interconnected} with each other and how they might coordinate their activities.

\subsection{Potential Implications}
There are several potential implications of the study reported here:
\begin{itemize}
\item {\em The promise of  Top-Down Approach:} Our results show that the still-unproven top-down approach has considerable merit, at least, in the case of transcriptional regulatory networks. The structural features of the analyzed networks seem to have significant functional implications, as verified using the GO database. We must note, however, that the evaluation of the functional significance of the partitioning procedure is limited by the fact that current knowledge about cellular systems is incomplete.  For example, {\em the network contains several hundred genes whose functionalities are unknown}. Moreover, even for genes whose functional descriptions are found in the GO database, it might be missing  key regulatory interactions that are nonetheless captured in the experimentally inferred network. Thus, the communities of genes and regulators and their connectivity structure {\em might hold much more functional information}  than what is revealed by the GO database. For example, we are currently investigating the differences and similarities in the topological characteristics of the different functional blocks.

\item {\em Tracking Context-Sensitive Reorganization of Functional Blocks:}  One can now apply our community partitioning analysis on the regulatory networks  derived {\em for the same organism} but when {\em cells are subjected to different conditions.} Regulatory networks obtained from such experiments have been already reported by the same group at MIT for {\em S.cerevisiae}, and they have also reported~\cite{harbison} some of the changes in the network connectivity  as a result of varying the environmental conditions, e.g., the genes regulated by certain regulators change considerably from one condition to another, and thus making fairly significant changes in the regulatory networks.  {\em We are currently studying the changes in the community structure} as the conditions are varied.  This could lead to a better understanding of how the functional blocks get reorganized  and how different blocks merge or get split as the organism reacts to different conditions.

\item {\em Comparisons Across Organisms:} Our results suggest a systematic means for exploring both the functional organization of an unstudied organism, and for comparing the community structures across organisms. For example, one could study how  the communities and their relationships change from species to species.  This could elucidate different organizations of functional blocks and their diversity and any evolutionary footprint that might be gleaned from analyzing the regulatory modules. Similarly, for an organism that has not been studied, a genome wide location analysis could be used to obtain its regulatory network. The communities in the regulatory networks can then be investigated for functional significance, using known instances of structure-vs-functional relationships observed in other organisms. This could lead to an automated and a faster means for deciphering salient characteristics and distinguishing features of the unstudied organism.

\item{\em A Pedagogical Shift?} Biochemistry text books have mostly followed the lead of the previously-discussed bottom-up approach to exploring cellular systems. Perhaps, an equally useful alternative would be to use large-scale  networks, as obtained from high-throughput experiments, as guides to naturally unfold a detailed description of the organization and architecture of cellular systems. Figs. 1 and 2, embellished with more detailed annotations, seem to be good candidates for what might be an introductory chapter of a systems biology textbook, and a guide to  how the different chapters (e.g., each corresponding to a community) might be organized and interlinked. Recall that {\em these annotated figures were generated} from a large-scale transcriptional regulatory network {\em in an automated fashion}. Such a network-based exposition provides a multi-dimensional view of the system, capturing the different functional blocks at different scales and  in different functional relationships with others.
\end{itemize}

\subsection{Notes on Previous Work}
Gene networks, in general, and transcriptional regulatory networks, in particular, have been studied for a while.  For example, several large-scale properties of transcription regulatory networks (including the particular network used in our study)  have been thoroughly investigated. It has been shown~\cite{guelzim-NG}\cite{babu-cosb} that the out-degree (the number of genes regulated by each factor) typically follows a power law, while the in-degree distribution (the number of regulators affecting the promoter region of each gene) has an exponential decay, thus demonstrating that these networks share several properties with scale-free topologies~\cite{eriksen}. In a separate effort, characteristic regulatory motifs have been identified~\cite{alon-science}, by comparing their frequency of occurrence with that of randomly generated topologies having the same large scale properties of transcriptional networks~\cite{shen-orr}. Typically, {\em these motifs involve a very small number of nodes} (2$\div$10) and are organized in a finite set of simple structures, such as single-input or multi-input regulatory modules, feed-forward loops and so on. These structures are very closely related to well-known transcriptional regulation units, such as operons and regulons. However, due to the size and complexity of the connectivity structure, the {\em subdivision of such networks into large, and frequently inter-connected, regulatory modules was not addressed}.

As for inferring functional blocks,  several methodologies have been devised that combine different types of high-throughput data sources in order to build functionally coherent modules of genes and transcriptional regulators~\cite{wyrick}. For example in~\cite{ziv-nature}, an algorithm is described that combines information from genome-wide location analysis and expression data sets in order to identify regulatory networks of gene modules, where the latter are defined as a set of genes which are both co-expressed and also share a common set of transcriptional factors that are known to bind to their promoter regions. An approach that assigns genes to context-dependent and potentially overlapping ``transcription modules'', is described in~\cite{ihmels-nature}. The method clusters co-regulated groups of genes based on their expression levels measured under specific experimental conditions. In~\cite{segal-NG}, the authors introduce a method which starting from a gene expression data set and a pre-compiled set of candidate transcriptional regulators, simultaneously identifies a partition of genes into modules, as well as a regulatory program, \emph{i.e.,} a set of rules that explains the expression behavior of the members of each module. Finally, an example of an approach based on integrating the analysis of common sequence motifs in genes' promoter regions with expression level data is described in~\cite{segal-recomb02}.

Lastly, we note that topology-based community finding techniques have been applied to biological data in at least a couple of examples. The first is the analysis of protein-protein interaction databases~\cite{spirin}. The second is the organization of literature data, where networks of gene co-occurrences are extracted by parsing the abstract of scientific articles covering a specific topic\cite{huberman}. However, the fact that the topology partitioning algorithms are capable of identifying well-defined functional units, as shown in our paper, is to our knowledge, the \emph{first compelling evidence of a significant association between structure and function} in cellular networks.

\subsection{Background on Community Finding Algorithms}
Network topologies that are rich in structure have been studied in several non-biological fields. One example is given by Internet browsing patterns, which can be effectively represented as directed graphs linking users with the sites they tend to visit most frequently. Another example is represented by peer-to-peer networks, where computer users are connected either by a file sharing architecture or, through a social network type of infrastructure. In both cases, methods have been devised that are capable of automatically partitioning the nodes in the network into groups or \emph{``communities''}. Typically, what characterizes a community is the fact that nodes within the community itself are densely inter-connected, while such connections tend to be sparser between members of different communities.

Several examples of network partitioning algorithms are described in the literature~\cite{newman-complex}. Among them are spectral bisection, the Kernighan Lin algorithm, and the Girvan and Newman's algorithm~\cite{newman-girvan}\cite{girvan-pnas}, just to cite a few. The latter is of particular interest having found application to several different types of networks, such as scientific collaboration networks, social networks, and the World Wide Web, with successful outcomes~\cite{newman-assort}. As we note in the Discussion section, the GN algorithm suffers from several drawbacks, and we are pursuing more flexible community partitioning approaches, which will be better equipped to capture the complexity of cellular systems.

\section{Methods}

The transcriptional network topology was derived from a whole-genome binding site location analysis of the yeast \emph{Saccharomyces Cerevisiae}~\cite{harbison}. The experimental procedure identifies the binding  affinities between a set of 203 transcriptional regulators and the yeast DNA, under different experimental conditions. With a high confidence level ($P < 0.001$) the promoter regions of a total of 2,845 genes were identified as targets of regulation, for a total of 6,170 regulatory interactions, in rich media conditions.

We analyzed the resulting network topology by using the faster implementation of Girvan and Newman~\cite{newman-fast} algorithm, which is particularly suitable for handling large networks involving several thousands of nodes. The algorithm is based on the idea of successively removing edges with the highest degree of "betweenness" until a final partitioning of the nodes is obtained. The degree of "betweenness" measures the likelihood that a particular edge lies between two separate communities: in~\cite{newman-girvan} this is calculated by finding the shortest path between any two nodes in the network and counting the frequency with which each edge is traversed. The edges that are traversed more often are likely to be interconnecting separate communities in the network. To determine the optimal number of edges to be removed, the algorithm relies on the notion of \emph{modularity} \emph{Q}\cite{newman-girvan}, which provides a measure of the fraction of within-community edges minus the expected value of the same quantity in a network with the same community partitioning but random connections between its nodes.

Values of $Q$ above 0.3 have been suggested~\cite{newman-girvan} as meaningful for identifying significant partitions. In order to choose a statistically significant level of the modularity index, we ran the community finding algorithm on 1,000 randomly generated topologies, characterized by the same connectivity pattern as the transcriptional network of \emph{S.cerevisiae}, but with the edges assigned at random. We found that a modularity threshold of $Q=0.5$ is sufficient to guarantee that the partitioned structures are significant and not simply due to general large-scale properties of the network (for $Q = 0.5$ no partitioning is found in any of the randomly generated topologies).

Once a top level partitioning is achieved, each of the resulting communities is evaluated for further subdivision. Because of the scale-free nature of this type of networks, communities tend to considerably vary in size (ranging between 3 and 258 nodes). Up to three nested levels of partitioning were considered.

Since the connectivity topology is determined through an experimental procedure~\cite{harbison}, {\em it is critical to determine how stable is the outcome of the partitioning algorithm}. Therefore, besides assessing the biological significance of the resulting community structure, we describe a procedure for evaluating the statistical robustness of the network partitioning results, based on systematically introducing random errors in the connectivity topology.

\section{Results}

\begin{table*}
\footnotesize
\centering
\vspace*{-5mm}
\begin{tabular}{c|c|c|c|c|l|l} \hline\hline
\mc{3}{c|}{ID} & \#TFs & \# genes & GO Annotation & p-value \\ \hline\hline
{\bf\small 0.0} & &   & 27 & 415 & {\bf\small Amino acid biosynthesis, transport; cell growth}\\
    & 0.0.0 &         & 8 & 96 & Amino acid biosynthesis/metabolism & {\bf\red 4.49e-22} \\
    &       & 0.0.0.0 & 1 & 39 & Amino acid biosynthesis (ornithine) & {\bf\red 4.67e-15} \\
    &       & 0.0.0.1 & 2 & 19 & Amino acid transport & {\bf\red 2.11e-08} \\
    &       & 0.0.0.2 & 2 & 14 & Arginine biosynthesis & {\bf\red 2.46e-06} \\
    &       & 0.0.0.3 & 1 & 14 & Branched chain family amino acid biosynthesis (leucine) & {\bf\red 3.59e-11} \\
    &       & 0.0.0.4 & 2 & 10 & Urea cycle intermediate biosynthesis & 6.75e-05 \\
    & 0.0.1 &         & 6 & 67 & Carbohydrate transport, sterol transport & {\bf\red 7.25e-10} \\
    & 0.0.2 &         & 1 & 55 & Carbohydrate metabolism & 3.05e-3 \\
    & 0.0.3 &         & 3 & 51 & Iron ion transport, hexose transport & 4.43e-3 \\
    & 0.0.4 &         & 2 & 46 & Response to metal ion & 8.17e-5 \\
    & 0.0.5 &         & 2 & 38 & Nucleoside and nucleotide metabolism & 6.44e-3 \\
    & 0.0.6 &         & 3 & 35 & Cell wall organization and biogenesis & 4.0e-4 \\
    & 0.0.7 &         & 2 & 27 & Gluconeogenesis, carbohydrate biosynthesis & 1.55e-3 \\  \hline
{\bf\small 0.1} & &   & 17 & 405 & {\bf\small Cell cycle regulation} \\
    & 0.1.0 &         & 6 & 138 & Cell cycle, DNA replication and chromosome cycle & {\bf\red 2.99e-7} \\
    & 0.1.1 &         & 3 & 96 & Cell cycle, M phase & 1.0e-4 \\
    & 0.1.2 &         & 3 & 81 & Cell cycle, intracellular transport & 1.4e-4 \\
    & 0.1.3 &         & 3 & 64 & Conjugation with cellular fusion, sexual reproduction & {\bf\red 1.05e-9} \\
    & 0.1.4 &         & 2 & 26 & Response to stimulus, regulation of cell cycle & 6.32e-3 \\ \hline
{\bf\small 0.2} & &   & 31 & 384 & {\bf\small DNA recombination, maintenance} \\
    & 0.2.0 &         & 6 & 102 & Cytokinesis, completion of separation & 8.79e-5 \\
    & 0.2.1 &         & 4 & 74 & Telomerase dependent telomere maintenance & {\bf\red 2.64e-7} \\
    & 0.2.2 &         & 8 & 60 & Establishment and/or maintenance of chromatin architecture & {\bf\red 1.10e-07} \\
    &       & 0.2.2.0 & 2 & 17 & Chromatin assembly/disassembly, DNA packaging & {\bf\red 3.12e-11} \\
    &       & 0.2.2.1 & 1 & 17 & Cytoplasm organization and biogenesis & 2.23e-2 \\
    &       & 0.2.2.2 & 2 & 9 & RNA processing & 6.18e-2\\
    &       & 0.2.2.3 & 2 & 8 & Unknown function  & n/a \\
    &       & 0.2.2.4 & 1 & 9 & Response to pheromone during conjugation & 1.45e-3 \\
    & 0.2.3 &         & 8 & 42 & Mitotic recombination & 2.49e-3\\
    & 0.2.4 &         & 2 & 43 & Cell organization and biogenesis & {\bf\red 9.97e-7}\\
    & 0.2.5 &         & 2 & 39 & ATP synthesis coupled proton transport & {\bf\red 1.17e-11} \\
    & 0.2.6 &         & 1 & 24 & Protein folding, response to stress & 7.3e-4 \\ \hline
{\bf\small 0.3} & & & 5 & 253 & {\bf\small Protein biosynthesis, RNAs metabolism} & {\bf\red 8.08e-9} \\\hline
{\bf\small 0.4} & &   & 25 & 231 & {\bf\small Protein synthesis, transport and glycosylation} \\
    & 0.4.0 &         & 3 & 38 & Cell cycle checkpoint & 2.32e-2 \\
    & 0.4.1 &         & 1 & 35 & Regulation of transcription & 5.54e-2 \\
    & 0.4.2 &         & 4 & 29 & Protein polyubiquitination & 2.46e-2 \\
    &       & 0.4.2.0 & 1 & 10 & RNA metabolism & 2.09e-2 \\
    &       & 0.4.2.1 & 1 & 8 & Cofactor metabolism & 8.08e-3 \\
    &       & 0.4.2.2 & 1 & 7 & Macromolecule catabolism & 1.43e-2 \\
    &       & 0.4.2.3 & 1 & 4 & Nucleic acid metabolism & 1.32e-3 \\
    & 0.4.3 &         & 4 & 27 & Regulation of protein biosynthesis & 3.91e-3 \\
    & 0.4.4 &         & 3 & 27 & Glucose metabolism & 4.1e-4 \\
    & 0.4.5 &         & 2 & 25 & Protein transport and localization (ER to Golgi) & 1.99e-2\\
    & 0.4.6 &         & 5 & 18 & Intracellular protein transport & 5e-4 \\
    & 0.4.7 &         & 2 & 18 & Protein amino acid glycosylation & 5.5e-4 \\
    & 0.4.8 &         & 1 & 14 & Golgi vesicle transport, protein localization & 9.35e-3 \\
\end{tabular}
\caption{List of functional modules obtained by partitioning the topology of the transcriptional network of \emph{S.cerevisiae}. The annotation for each community was obtained from the Gene Ontology database. The significance of the enrichment is expressed by the \emph{p-value} associated with the list of nodes in each module (with the most significant highlighted in red).}
\label{yeast-com-table-1}
\end{table*}

\begin{table*}
\footnotesize
\centering
\begin{tabular}{c|c|c|c|l|l} \hline\hline
\mc{2}{c|}{ID} & \#TFs & \# genes & GO Annotation & p-value \\ \hline\hline
{\bf\small 0.5}  &        & 8 & 231 & {\bf\small Protein catabolism, secretory pathway} \\
     & 0.5.0  & 1 & 113 & Ubiquitin-dependent protein catabolism, peptidolysis & 1.5e-4 \\
     & 0.5.1  & 1 & 48 & Sporulation, spore wall assembly & {\bf\red 7.25e-7} \\
     & 0.5.2  & 1 & 26 & mRNA catabolism, deadenylation-dependent decay & 2.58e-3 \\
     & 0.5.3  & 1 & 19 & Secretory pathway & 2.59e-3 \\
     & 0.5.4  & 1 & 14 & G1-specific transcription in mitotic cell cycle & 3.3e-4 \\
     & 0.5.5  & 3 & 11 & SRP-dependent protein-membrane targeting, translocation & 1.0e-4 \\ \hline
{\bf\small 0.6}  & & 4 & 189 & {\bf\small Ribosomes biogenesis and assembly} & {\bf\red 6.59e-12} \\ \hline
{\bf\small 0.7} & & 9 & 177 & {\bf\small Lipid biosynthesis, degradation} \\
     & 0.7.0  & 1 & 86 & Ergosterol metabolism, NADH metabolism & {\bf\red 4.12e-13} \\
     & 0.7.1  & 3 & 50 & Lipid biosynthesis, phospholipid metabolism & {\bf\red 1.08e-12} \\
     & 0.7.2  & 2 & 21 & ATP synthesis coupled electron transport & 1.22e-5 \\
     & 0.7.3  & 1 & 11 & Unknown & n/a \\
     & 0.7.4  & 1 & 6 & Proline, glutamate metabolism & {\bf\red 2.54e-6} \\
     & 0.7.5  & 1 & 3 & Unknown & n/a \\ \hline
{\bf\small 0.8} & & 13 & 131 & {\bf\small Aerobic respiration} \\
     & 0.8.0  & 3 & 48 & Ty element transposition, alcohol metabolism & 4e-4 \\
     & 0.8.1  & 2 & 39 & Organelle organization and biogenesis & 1.83e-3 \\
     & 0.8.2  & 3 & 32 & Sulfur metabolism, methionine metabolism & {\bf\red 2.81e-6} \\
     & 0.8.3  & 1 & 25 & Purine nucleotide metabolism & {\bf\red 7.66e-8} \\
     & 0.8.4  & 1 & 12 & Carboxylic acid metabolism & 6.83e-2 \\
     & 0.8.5  & 2 & 9 & Galactose metabolism & {\bf\red 1.03e-13} \\
     & 0.8.6  & 1 & 6 & Unknown & n/a \\ \hline
{\bf\small 0.9}  & & 5 & 136 & {\bf\small TCA cycle} & 8.92e-3 \\ \hline
{\bf\small 0.10} & & 6 & 100 & {\bf\small Response to chemical substance, drug response} \\
     & 0.10.0 & 1 & 50 & Response to chemical substance, drug transport & 8.54e-5 \\
     & 0.10.1 & 1 & 26 & Protein-vacuolar targeting & 9.3e-4 \\
     & 0.10.2 & 2 & 9 & DNA replication and chromosome cycle & 2.38e-3 \\
     & 0.10.3 & 1 & 9 & Unknown & n/a \\
     & 0.10.4 & 1 & 6 & Response to stress stimulus & 4.42e-3 \\ \hline
{\bf\small 0.11} & & 8 & 88 & {\bf\small Response to oxidative stress}\\
     & 0.11.0 & 2 & 27 & DNA replication initiation, chromatin remodeling & 9.2e-4  \\
     & 0.11.1 & 1 & 15 & Negative regulation of transcription from Pol II promoter & 3.17e-3 \\
     & 0.11.2 & 1 & 14 & Organelle organization and biogenesis & 6.58e-2 \\
     & 0.11.3 & 1 & 13 & Response to oxidative stress, cofactor biosynthesis & 2.72e-3  \\
     & 0.11.4 & 1 & 11 & Aromatic compound metabolism & 2.79e-3 \\
     & 0.11.5 & 1 & 6 & Unknown & n/a \\
     & 0.11.6 & 1 & 2 & Unknown & n/a \\ \hline
{\bf\small 0.12} & & 3 & 38 & {\bf\small Unknown}\\
     & 0.12.0 & 1 & 21 & Formate metabolism & {\bf\red 9.27e-8} \\
     & 0.12.1 & 1 & 11 & Transcription, DNA dependent & 2.45e-2 \\
     & 0.12.2 & 1 & 6 & Regulation of transcription, DNA dependent & 2.23e-2 \\ \hline
{\bf\small 0.13} & & 1 & 9 & {\bf\small Coenzyme metabolism, cofactor metabolism} & 6.6e-3 \\ \hline
{\bf\small 0.14} & & 1 & 6 & {\bf\small Homeostasis} & 2.1e-3 \\ \hline
\end{tabular}
\caption{List of functional modules (\emph{cont.}) obtained by partitioning the topology of the transcriptional network of \emph{S.cerevisiae}. The annotation for each community was obtained from the Gene Ontology database. The significance of the enrichment is expressed by the \emph{p-value} associated with the list of nodes in each module (with the most significant highlighted in red).}
\label{yeast-com-table-2}
\end{table*}

Our procedure identified 15 top-level communities ($Q=0.6285$) that were further subdivided into a total of 79 modules, during the recursive stage (Fig.\ref{com-tree-struct} shows the tree structure of the resulting communities). The sensitivity of the resulting community structure to errors in the connectivity topology was evaluated by randomly adding or deleting a fraction (varying between 1\% and 20\%) of the edges and re-partitioning the graph according to the modified topology. Our results show that for a {\em range of false-positive and false-negative connections} similar to those reported for this experimental data set~\cite{harbison}, {\em the outcome of the partitioning algorithm is very consistent}. Although a small percentage of nodes are re-distributed across different modules, the community structure is mostly preserved.

Tables~\ref{yeast-com-table-1} and~\ref{yeast-com-table-2} provide a description of each community, where the annotation and the significance of the enrichment (measured as a p-value) were obtained from the Gene Ontology (GO) database~\cite{go-ref}. The number of genes and transcription factors present in  each module are also included. Each community is labeled according to its nesting level: for example modules labeled \emph{\{0.x.y.z\}} are sub-communities of \emph{\{0.x.y\}}, which are in turn sub-communities of \emph{\{0.x\}}. Community \emph{\{0\}} is the root module including all genes and regulators in the network.

\begin{figure*}[p]
\centerline{\includegraphics[width=7in,height=9in]{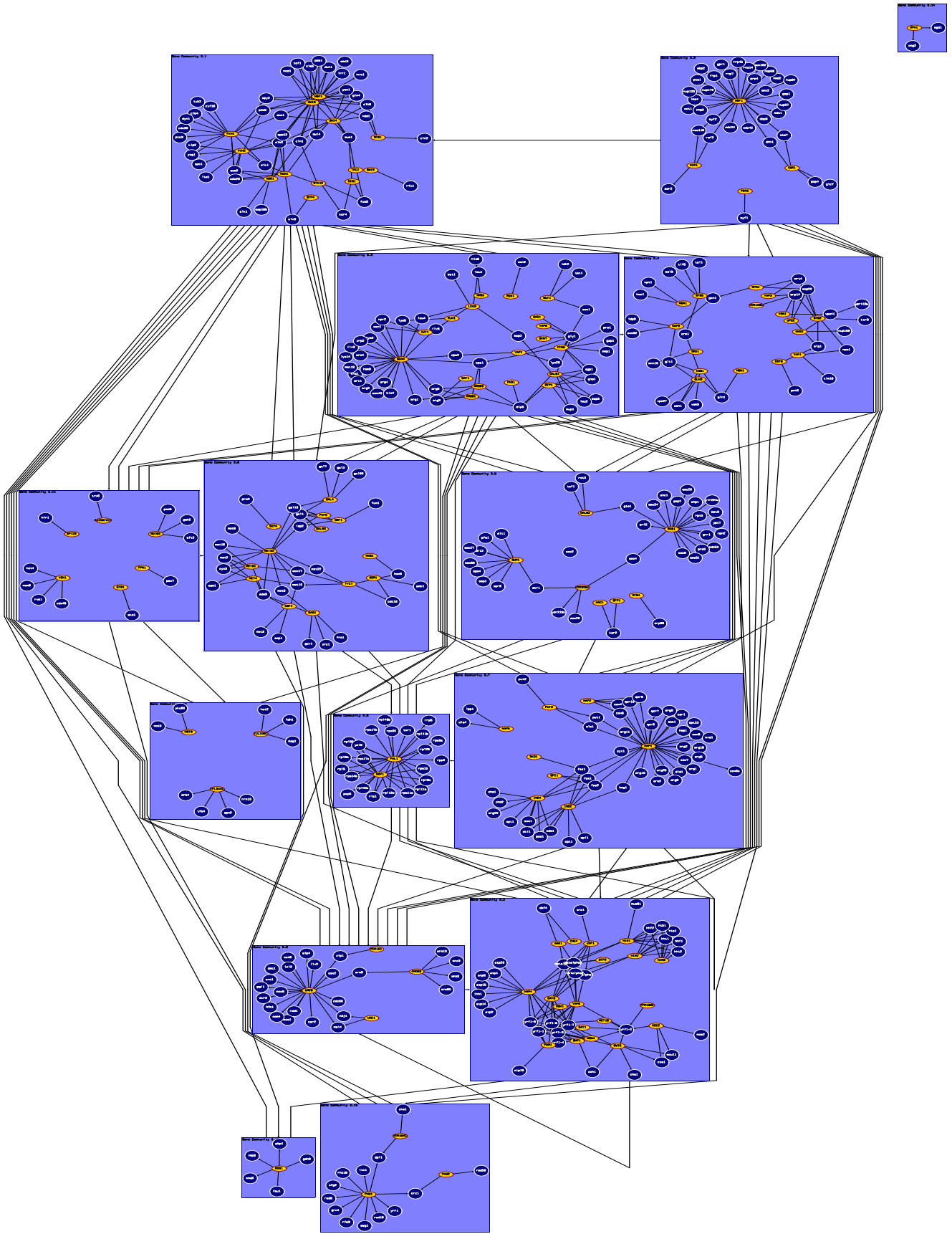}}
\caption{Global view of the organization of the various regulatory modules identified with the topology partitioning algorithm. Only the most relevant genes and regulators are shown in each module.}
\label{all-coms-graph}
\end{figure*}

The following paragraphs provide a description of the top-level communities identified with the procedure. Fig.~\ref{all-coms-graph} provides a global view of the organization structure of the regulatory modules.

\subsection{Sample communities}

The amino-acid biosynthesis module (Fig.~\ref{com-tree-struct}a) is the largest of the top-level communities involving a total of 417 genes and 25 TRs. The module includes a well-defined number of separate sub-structures associated to the synthesis of specific amino-acids (arginine, ornithine, glutamine and leucine), to amino-acid transport, and to the biosynthesis of intermediate products in the urea cycle. The community also includes several sub-modules related to carbohydrate transport and metabolism, a result consistent with the known relationship between carbon source pathways and amino acid synthesis.

Several cell division cycle genes (\emph{cdc5, cdc6, cdc7, cdc20, cdc39, cdc48}) are among the key members of community \emph{0.1} (Fig.~\ref{com-tree-struct}b) which is essentially associated to cell cycle related activities, including DNA replication and chromosome cycle as well as the regulation of the different stages of mitosis. Both G1 cyclins (\emph{cln1, cln2}, and \emph{cln3}) and B-type cyclins (\emph{clb2, clb4, clb5}, and \emph{clb6}), involved in activation of S, G2, and M phases of the cell cycle, are included among the genes in this module, together with several key players of G1/S transition (\emph{swe1, sap185}) and G2/M transition (\emph{Swi4, ndd1}). The partition also includes a separate sub-community implicated in conjugation with cellular fusion and sexual reproduction, which includes the genes \emph{mdg1, afr1} and \emph{mfa1} (all involved in signal transduction during conjugation), \emph{jem1, scw10, fus1} and \emph{fus3} (both regulated by mating pheromone and proposed to coordinate events required for fusion), \emph{gpa1} and the transcription factor \emph{Ste12} (activated by a MAPK signaling cascade), inducer of genes involved in mating.

A particularly well-characterized community is the one implicated in DNA recombination and maintenance (Fig.~\ref{com-tree-struct}c), encompassing specific sub-modules related to the establishment and maintenance of chromatin architecture (\emph{hhf1, hhf2, hht1, hht2, hta1, htb1}), telomere maintenance (\emph{yrf1--7}), and mating type specific regulators (\emph{Ash1}, \emph{Hml}$\alpha_1$, \emph{Hml}$\alpha_2$, \emph{Mat}$\alpha_1$, \emph{Mat}$\alpha_2$). This large community (381 genes), also includes two large sub-networks: the first is involved in nucleoside phosphate metabolism (\emph{atp1, atp2, atp5, atp14, atp15, atp19,} and \emph{atp20}), a precursor pathway of DNA molecule biosynthesis, while the second is linked to cytokinesis and completion of separation (\emph{scw11, cts1, chs1}).

Community \emph{0.3} (Fig.~\ref{com-tree-struct}d) comprises almost exclusively genes involved in different stages of protein biosynthesis and RNA processing, including pre-\emph{m}RNA splicing (\emph{smx2, syf2, cwc2, prp4, syf1}), polyadenylation (\emph{pta1, ref2, fip1, rna14}), and capping (\emph{cet1}), \emph{r}RNA processing (\emph{utp8, utp14, rrp7, rrp12, rrp45, nhp2, pop4}), and transport of the different RNAs (\emph{nup57, nup84, nup170, nup133, gbp2, pom152}). This module also includes several proteins which are implicated in vesicle-mediated transport, such as \emph{emp47, erv46, sec18, rer1, cop1, ykt6, erv25} (Endoplasmic Reticulum to Golgi transport),  \emph{vps33, pik1, sso1} (Golgi to endosome or plasma membrane transport).

Several important steps of protein metabolism in the cell can be assigned to different sub-structures of module \emph{0.4} (Fig.~\ref{com-tree-struct}e). In particular, the community includes genes associated to protein transport and localization, such as \emph{stp22, vps27, vps41, vps61, vps66, vps74} (protein vacuolar targeting), \emph{pam16} (mitochondrial matrix protein import), \emph{nup159} (\emph{m}RNA-nucleus export), \emph{atg7} and \emph{atg9} (protein autophagy), \emph{srp14} (protein-ER targeting), or \emph{tim18} (protein-membrane targeting). Separate sub-communities are involved in protein amino-acid glycosylation (\emph{ktr3, hoc1, alg1, mnn10, pmt5}) and protein ubiquitination (\emph{ubc11, ufd4}).

Closely related to the protein biosynthesis community is the module related to protein catabolism and the secretory pathway (Fig.~\ref{com-tree-struct}f). The largest of its different sub-structures (115 genes) is  involved in ubiquitin-dependent protein catabolism, proteolysis and peptidolysis (\emph{rpt5, shp1, rpn2, pre3, grr1, asi3}). Other relevant sub-modules are those related to the secretory pathway (\emph{sec4, sec8, sec20, sec23, sec24, sec27, snc1, ypt7}) and sporulation (\emph{spr3, pfs1, ssp1, dit1, dit2, sps4, dtr1, gts1}).

A module entirely dedicated to ribosome biogenesis and assembly is shown in Fig.\ref{com-tree-struct}g. This community involves the vast majority of ribosomal proteins as well as all the genes involved in the assembly of the large and small ribosomal subunits. Community \emph{0.7} is functionally associated to processes related to lipid metabolism, with several sub-modules spanning from lipid biosynthesis to lipid degradation in aerobic conditions.

Community \emph{0.8} and \emph{0.9} are both associated to pathways which are active in aerobic conditions, with a significant sub-component involved in galactose metabolism. Finally, community \emph{0.10} and \emph{0.11} are both highly specialized. The first includes several genes implicated in the response to chemical substances, drug transport, and response to DNA damage stimuli. The second is linked to oxidative stress response.

\subsection{Inter-community connectivity patterns}

\begin{figure*}[p]
\centerline{\includegraphics[width=7.5in]{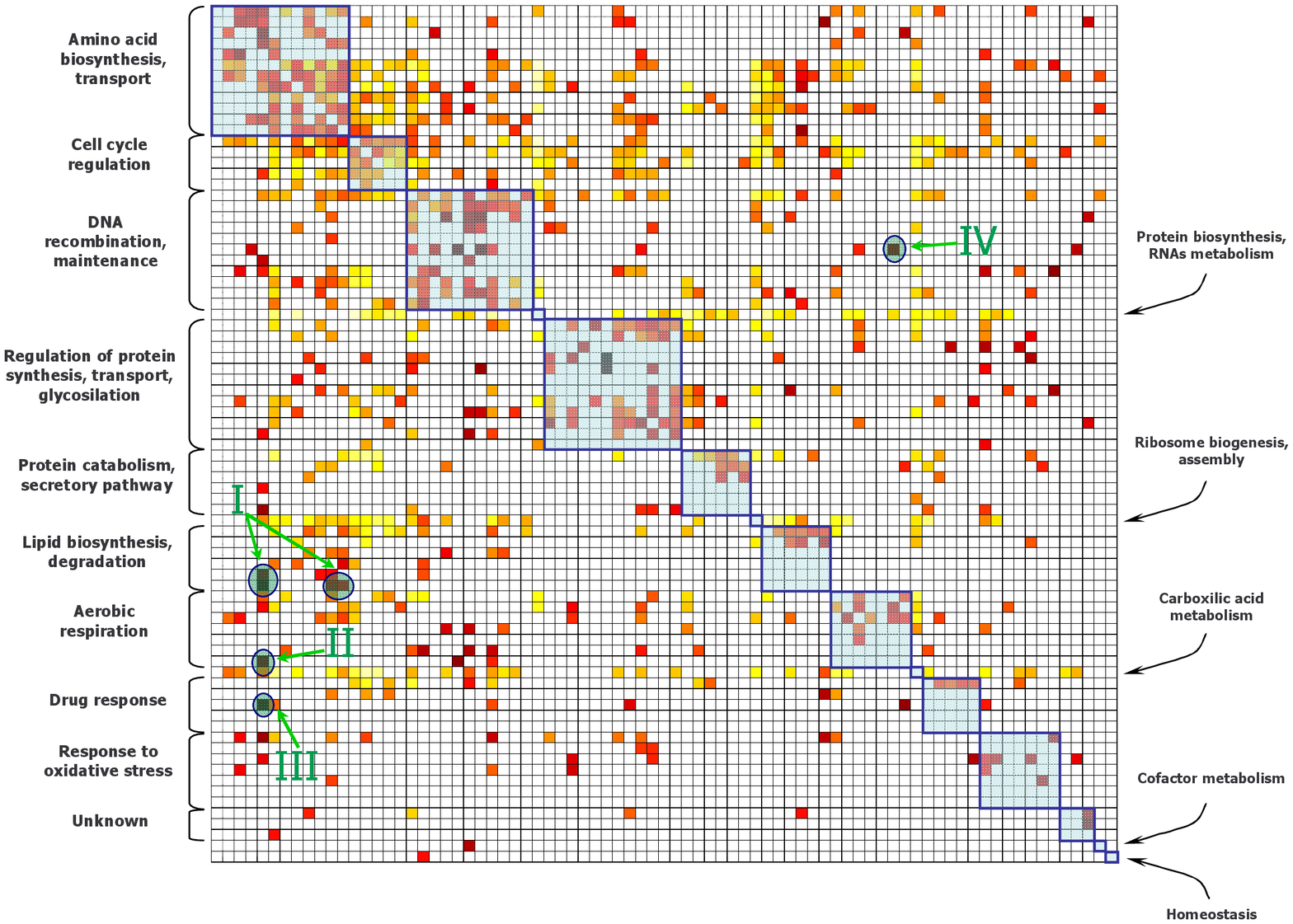}}
\caption{Map of the density of inter-community links. The darkest spots indicate a strong connection between two regulatory sub-modules, while the large shadowed boxes enclose the connectivity strengths among the 15 top-level communities. A set of relevant interactions are highlighted: strong co-regulation patterns appear between the the amino acid biosynthesis module and the lipid metabolism module \emph{(I)}, the aerobic respiration pathway \emph{(II)}, and the drug response system \emph{(III)}. The regulation of genes involved in resistance to arsenic compounds is behind the co-regulation pattern between the DNA maintenance module and the aerobic respiration module shown in \emph{(IV)}.}
\label{inter-conn-map}
\end{figure*}

Once a partitioning into communities is obtained, one can study how different types of regulatory modules are connected among each other. The relative density of edges running across the various communities provides an indication of the degree of co-regulation among members of different communities. Fig.~\ref{inter-conn-map}, shows a map of such density of inter-connections, with the large shadowed boxes enclosing the patterns of connectivity within each top-level community. A close examination of the highest edge densities reveals a number of significant patterns of connectivity among the discovered modules. Several of them associate the largest regulatory sub-network (amino acid biosynthesis, Fig.~\ref{com-tree-struct}a), with modules implicated in lipid metabolism and ATP synthesis coupled electron transport, in agreement with the known dependencies among these pathways. A detailed description of several highlighted co-regulation patterns is provided in the caption of Fig.~\ref{inter-conn-map}.

\section{Discussion}

Although the vast majority of the transcriptional sub-networks identified by the partitioning algorithm showed a significant functional coherence, we also found several examples of (often smaller) modules whose functional pertinence could not be easily determined (\emph{e.g.} communities \emph{0.7.3}, \emph{0.7.5}, \emph{0.8.6}, \emph{0.11.5}, and \emph{0.11.6}). In other cases, even when a general functional category could be assigned to a sub-module, the enrichment level of the resulting annotation was not deemed statistically significant (\emph{cfr.} Tables~\ref{yeast-com-table-1} and~\ref{yeast-com-table-2}). At least two factors limit our ability of evaluating the results of the partitioning procedure: the first is the limited amount of information currently available in ontology databases. The experimental procedure used to obtain the connectivity data involves several hundred genes whose molecular function or related biological process is unknown. Moreover, even in those cases when a functional description is available for the associated nodes, the experimental procedure is likely to capture regulatory interactions that have not been previously observed. The presence of a non-negligible amount of false-positives and false-negatives in the data is also likely to affect the outcome of the module partitioning procedure.

The method we employed for partitioning \emph{S.cerevisiae}'s transcriptional network topology can be improved in several ways. The most limiting aspect of the current procedure is that nodes cannot be assigned simultaneously to multiple modules. When testing the robustness of the community finding procedure against errors in the connectivity topology we discovered that the nodes that are more likely to be re-assigned to a different module are those that are weekly connected to the original module. Typically, these nodes play a role of links among different sub-networks and should not be uniquely assigned to a single partition. A limitation both of the data currently available and of the partitioning algorithm is the inability of accounting for the stochastic nature of the connectivity topology. A more robust framework would be one where regulatory interactions are assigned a probability density function and nodes are assigned to modules according to a measure of likelihood.

\bibliographystyle{apsrev}
\bibliography{arxiv-sub}

\end{document}

%% file: macros.tex
\newcommand{\bea}{\begin{eqnarray}}
\newcommand{\eea}{\end{eqnarray}}
\newcommand{\beqn}{\begin{equation}}
\newcommand{\eeqn}{\end{equation}}

\newcommand{\mc}[3]{\multicolumn{#1}{#2}{#3}}
\newcommand{\bt}[2]{\begin{table}[#2]  \begin{center} \begin{tabular}{#1}}
\newcommand{\et}[2]{\end{tabular} \end{center} \caption{#2} \label{#1}\end{table}}


%% file: arxiv-sub.bbl
\begin{thebibliography}{25}
\expandafter\ifx\csname natexlab\endcsname\relax\def\natexlab#1{#1}\fi
\expandafter\ifx\csname bibnamefont\endcsname\relax
  \def\bibnamefont#1{#1}\fi
\expandafter\ifx\csname bibfnamefont\endcsname\relax
  \def\bibfnamefont#1{#1}\fi
\expandafter\ifx\csname citenamefont\endcsname\relax
  \def\citenamefont#1{#1}\fi
\expandafter\ifx\csname url\endcsname\relax
  \def\url#1{\texttt{#1}}\fi
\expandafter\ifx\csname urlprefix\endcsname\relax\def\urlprefix{URL }\fi
\providecommand{\bibinfo}[2]{#2}
\providecommand{\eprint}[2][]{\url{#2}}

\bibitem[{\citenamefont{Venter et~al.}(1996)\citenamefont{Venter, Smith, and
  Hood}}]{venter-nature}
\bibinfo{author}{\bibfnamefont{J.~C.} \bibnamefont{Venter}},
  \bibinfo{author}{\bibfnamefont{H.~O.} \bibnamefont{Smith}}, \bibnamefont{and}
  \bibinfo{author}{\bibfnamefont{L.}~\bibnamefont{Hood}},
  \bibinfo{journal}{Nature} \textbf{\bibinfo{volume}{381}},
  \bibinfo{pages}{364} (\bibinfo{year}{1996}).

\bibitem[{\citenamefont{Venter et~al.}(1998)}]{venter-science}
\bibinfo{author}{\bibfnamefont{J.~C.} \bibnamefont{Venter}}
  \bibnamefont{et~al.}, \bibinfo{journal}{Science}
  \textbf{\bibinfo{volume}{280}}, \bibinfo{pages}{1540} (\bibinfo{year}{1998}).

\bibitem[{\citenamefont{Ihmels et~al.}(2002)\citenamefont{Ihmels, Friedlander,
  Bergmann, Sarig, Ziv, and Barkai}}]{ihmels-nature}
\bibinfo{author}{\bibfnamefont{J.}~\bibnamefont{Ihmels}},
  \bibinfo{author}{\bibfnamefont{G.}~\bibnamefont{Friedlander}},
  \bibinfo{author}{\bibfnamefont{S.}~\bibnamefont{Bergmann}},
  \bibinfo{author}{\bibfnamefont{O.}~\bibnamefont{Sarig}},
  \bibinfo{author}{\bibfnamefont{Y.}~\bibnamefont{Ziv}}, \bibnamefont{and}
  \bibinfo{author}{\bibfnamefont{N.}~\bibnamefont{Barkai}},
  \bibinfo{journal}{Nature Genetics} \textbf{\bibinfo{volume}{31}},
  \bibinfo{pages}{370} (\bibinfo{year}{2002}).

\bibitem[{\citenamefont{Segal et~al.}(2003)\citenamefont{Segal, Shapira, Regev,
  Pe'er, Botstein, Koller, and Friedman}}]{segal-NG}
\bibinfo{author}{\bibfnamefont{E.}~\bibnamefont{Segal}},
  \bibinfo{author}{\bibfnamefont{M.}~\bibnamefont{Shapira}},
  \bibinfo{author}{\bibfnamefont{A.}~\bibnamefont{Regev}},
  \bibinfo{author}{\bibfnamefont{D.}~\bibnamefont{Pe'er}},
  \bibinfo{author}{\bibfnamefont{D.}~\bibnamefont{Botstein}},
  \bibinfo{author}{\bibfnamefont{D.}~\bibnamefont{Koller}}, \bibnamefont{and}
  \bibinfo{author}{\bibfnamefont{N.}~\bibnamefont{Friedman}},
  \bibinfo{journal}{Nature Genetics} \textbf{\bibinfo{volume}{34}},
  \bibinfo{pages}{166} (\bibinfo{year}{2003}).

\bibitem[{\citenamefont{Hughes et~al.}(2000)}]{hughes-rosetta}
\bibinfo{author}{\bibfnamefont{T.~R.} \bibnamefont{Hughes}}
  \bibnamefont{et~al.}, \bibinfo{journal}{Cell} \textbf{\bibinfo{volume}{102}},
  \bibinfo{pages}{109} (\bibinfo{year}{2000}).

\bibitem[{\citenamefont{Lee et~al.}(2002)}]{lee-science}
\bibinfo{author}{\bibfnamefont{T.~I.} \bibnamefont{Lee}} \bibnamefont{et~al.},
  \bibinfo{journal}{Science} \textbf{\bibinfo{volume}{298}},
  \bibinfo{pages}{799} (\bibinfo{year}{2002}).

\bibitem[{\citenamefont{Liao et~al.}(2003)\citenamefont{Liao, Boscolo, Yang,
  Tran, Sabatti, and Roychowdhury}}]{pnas-nca}
\bibinfo{author}{\bibfnamefont{J.~C.} \bibnamefont{Liao}},
  \bibinfo{author}{\bibfnamefont{R.}~\bibnamefont{Boscolo}},
  \bibinfo{author}{\bibfnamefont{Y.-L.} \bibnamefont{Yang}},
  \bibinfo{author}{\bibfnamefont{L.~M.} \bibnamefont{Tran}},
  \bibinfo{author}{\bibfnamefont{C.}~\bibnamefont{Sabatti}}, \bibnamefont{and}
  \bibinfo{author}{\bibfnamefont{V.~P.} \bibnamefont{Roychowdhury}},
  \bibinfo{journal}{Proceedings of the National Academy of Sciences (PNAS)}
  \textbf{\bibinfo{volume}{100}}, \bibinfo{pages}{15522}
  (\bibinfo{year}{2003}).

\bibitem[{\citenamefont{Gardner et~al.}(2003)\citenamefont{Gardner,
  di~Bernardo, Lorenz, and Collins}}]{gardner-science}
\bibinfo{author}{\bibfnamefont{T.~S.} \bibnamefont{Gardner}},
  \bibinfo{author}{\bibfnamefont{D.}~\bibnamefont{di~Bernardo}},
  \bibinfo{author}{\bibfnamefont{D.}~\bibnamefont{Lorenz}}, \bibnamefont{and}
  \bibinfo{author}{\bibfnamefont{J.~J.} \bibnamefont{Collins}},
  \bibinfo{journal}{Science} \textbf{\bibinfo{volume}{301}},
  \bibinfo{pages}{102} (\bibinfo{year}{2003}).

\bibitem[{\citenamefont{Newman and Girvan}(2004)}]{newman-girvan}
\bibinfo{author}{\bibfnamefont{M.~E.} \bibnamefont{Newman}} \bibnamefont{and}
  \bibinfo{author}{\bibfnamefont{M.}~\bibnamefont{Girvan}},
  \bibinfo{journal}{Physical Review E (Statistical, Nonlinear, and Soft Matter
  Physics)} \textbf{\bibinfo{volume}{69}}, \bibinfo{pages}{026113}
  (\bibinfo{year}{2004}).

\bibitem[{\citenamefont{Harris et~al.}(2004)}]{go-ref}
\bibinfo{author}{\bibfnamefont{M.~A.} \bibnamefont{Harris}}
  \bibnamefont{et~al.}, \bibinfo{journal}{Nucleic Acids Res.}
  \textbf{\bibinfo{volume}{32}}, \bibinfo{pages}{D258} (\bibinfo{year}{2004}).

\bibitem[{\citenamefont{Harbison et~al.}(2004)}]{harbison}
\bibinfo{author}{\bibfnamefont{C.~T.} \bibnamefont{Harbison}}
  \bibnamefont{et~al.}, \bibinfo{journal}{Nature}
  \textbf{\bibinfo{volume}{431}}, \bibinfo{pages}{99} (\bibinfo{year}{2004}).

\bibitem[{\citenamefont{Guelzim et~al.}(2002)\citenamefont{Guelzim, Bottani,
  Bourgine, and Kepes}}]{guelzim-NG}
\bibinfo{author}{\bibfnamefont{N.}~\bibnamefont{Guelzim}},
  \bibinfo{author}{\bibfnamefont{S.}~\bibnamefont{Bottani}},
  \bibinfo{author}{\bibfnamefont{P.}~\bibnamefont{Bourgine}}, \bibnamefont{and}
  \bibinfo{author}{\bibfnamefont{F.}~\bibnamefont{Kepes}},
  \bibinfo{journal}{Nature Genetics} \textbf{\bibinfo{volume}{31}},
  \bibinfo{pages}{60} (\bibinfo{year}{2002}).

\bibitem[{\citenamefont{Babu et~al.}(2004)\citenamefont{Babu, Luscombe,
  Aravind, Gerstein, and Teichmann}}]{babu-cosb}
\bibinfo{author}{\bibfnamefont{M.~M.} \bibnamefont{Babu}},
  \bibinfo{author}{\bibfnamefont{N.~M.} \bibnamefont{Luscombe}},
  \bibinfo{author}{\bibfnamefont{L.}~\bibnamefont{Aravind}},
  \bibinfo{author}{\bibfnamefont{M.}~\bibnamefont{Gerstein}}, \bibnamefont{and}
  \bibinfo{author}{\bibfnamefont{S.~A.} \bibnamefont{Teichmann}},
  \bibinfo{journal}{Current Opinion in Structural Biology}
  \textbf{\bibinfo{volume}{14}}, \bibinfo{pages}{283–} (\bibinfo{year}{2004}).

\bibitem[{\citenamefont{Eriksen and {H\"ornquist}}(2001)}]{eriksen}
\bibinfo{author}{\bibfnamefont{K.~A.} \bibnamefont{Eriksen}} \bibnamefont{and}
  \bibinfo{author}{\bibfnamefont{M.}~\bibnamefont{{H\"ornquist}}},
  \bibinfo{journal}{Physical Review E (Statistical, Nonlinear, and Soft Matter
  Physics)} \textbf{\bibinfo{volume}{65}}, \bibinfo{pages}{017102}
  (\bibinfo{year}{2001}).

\bibitem[{\citenamefont{Milo et~al.}(2002)\citenamefont{Milo, Shen-Orr,
  Itzkovitz, Kashtan, Chklovskii, and Alon}}]{alon-science}
\bibinfo{author}{\bibfnamefont{R.}~\bibnamefont{Milo}},
  \bibinfo{author}{\bibfnamefont{S.}~\bibnamefont{Shen-Orr}},
  \bibinfo{author}{\bibfnamefont{S.}~\bibnamefont{Itzkovitz}},
  \bibinfo{author}{\bibfnamefont{N.}~\bibnamefont{Kashtan}},
  \bibinfo{author}{\bibfnamefont{D.}~\bibnamefont{Chklovskii}},
  \bibnamefont{and} \bibinfo{author}{\bibfnamefont{U.}~\bibnamefont{Alon}},
  \bibinfo{journal}{Science} \textbf{\bibinfo{volume}{298}},
  \bibinfo{pages}{824} (\bibinfo{year}{2002}).

\bibitem[{\citenamefont{Shen-Orr et~al.}(2002)\citenamefont{Shen-Orr, Milo,
  Mangan, and Alon}}]{shen-orr}
\bibinfo{author}{\bibfnamefont{S.~S.} \bibnamefont{Shen-Orr}},
  \bibinfo{author}{\bibfnamefont{R.}~\bibnamefont{Milo}},
  \bibinfo{author}{\bibfnamefont{S.}~\bibnamefont{Mangan}}, \bibnamefont{and}
  \bibinfo{author}{\bibfnamefont{U.}~\bibnamefont{Alon}},
  \bibinfo{journal}{Nature Genetics} \textbf{\bibinfo{volume}{31}},
  \bibinfo{pages}{64} (\bibinfo{year}{2002}).

\bibitem[{\citenamefont{Wyrick and Young}(2002)}]{wyrick}
\bibinfo{author}{\bibfnamefont{J.~J.} \bibnamefont{Wyrick}} \bibnamefont{and}
  \bibinfo{author}{\bibfnamefont{R.~A.} \bibnamefont{Young}},
  \bibinfo{journal}{Current Opinion in Genetics \& Development}
  \textbf{\bibinfo{volume}{12}}, \bibinfo{pages}{130} (\bibinfo{year}{2002}).

\bibitem[{\citenamefont{Bar-Joseph et~al.}(2003)}]{ziv-nature}
\bibinfo{author}{\bibfnamefont{Z.}~\bibnamefont{Bar-Joseph}}
  \bibnamefont{et~al.}, \bibinfo{journal}{Nature Biotechnology}
  \textbf{\bibinfo{volume}{21}}, \bibinfo{pages}{1337} (\bibinfo{year}{2003}).

\bibitem[{\citenamefont{Segal et~al.}(2002)\citenamefont{Segal, Barash, Simon,
  Friedman, and Koller}}]{segal-recomb02}
\bibinfo{author}{\bibfnamefont{E.}~\bibnamefont{Segal}},
  \bibinfo{author}{\bibfnamefont{Y.}~\bibnamefont{Barash}},
  \bibinfo{author}{\bibfnamefont{I.}~\bibnamefont{Simon}},
  \bibinfo{author}{\bibfnamefont{N.}~\bibnamefont{Friedman}}, \bibnamefont{and}
  \bibinfo{author}{\bibfnamefont{D.}~\bibnamefont{Koller}}, in
  \emph{\bibinfo{booktitle}{RECOMB '02: Proceedings of the sixth annual
  international conference on Computational biology}} (\bibinfo{publisher}{ACM
  Press}, \bibinfo{year}{2002}), pp. \bibinfo{pages}{263--272}, ISBN
  \bibinfo{isbn}{1-58113-498-3}.

\bibitem[{\citenamefont{Spirin and Mirny}(2003)}]{spirin}
\bibinfo{author}{\bibfnamefont{V.}~\bibnamefont{Spirin}} \bibnamefont{and}
  \bibinfo{author}{\bibfnamefont{L.~A.} \bibnamefont{Mirny}},
  \bibinfo{journal}{Proc. Natl. Acad. Sci. USA} \textbf{\bibinfo{volume}{100}},
  \bibinfo{pages}{12123–} (\bibinfo{year}{2003}).

\bibitem[{\citenamefont{Wilkinson and Huberman}(2004)}]{huberman}
\bibinfo{author}{\bibfnamefont{D.~M.} \bibnamefont{Wilkinson}}
  \bibnamefont{and} \bibinfo{author}{\bibfnamefont{B.~A.}
  \bibnamefont{Huberman}}, \bibinfo{journal}{Proc. Natl. Acad. Sci. USA}
  \textbf{\bibinfo{volume}{101}}, \bibinfo{pages}{5241–}
  (\bibinfo{year}{2004}).

\bibitem[{\citenamefont{Newman}(2003)}]{newman-complex}
\bibinfo{author}{\bibfnamefont{M.~E.} \bibnamefont{Newman}},
  \bibinfo{journal}{SIAM Review} \textbf{\bibinfo{volume}{45}},
  \bibinfo{pages}{167} (\bibinfo{year}{2003}).

\bibitem[{\citenamefont{Girvan and Newman}(2002)}]{girvan-pnas}
\bibinfo{author}{\bibfnamefont{M.}~\bibnamefont{Girvan}} \bibnamefont{and}
  \bibinfo{author}{\bibfnamefont{M.~E.} \bibnamefont{Newman}},
  \bibinfo{journal}{Proc. Natl. Acad. Sci. USA} \textbf{\bibinfo{volume}{99}},
  \bibinfo{pages}{7821} (\bibinfo{year}{2002}).

\bibitem[{\citenamefont{Newman and Girvan}(2003)}]{newman-assort}
\bibinfo{author}{\bibfnamefont{M.~E.} \bibnamefont{Newman}} \bibnamefont{and}
  \bibinfo{author}{\bibfnamefont{M.}~\bibnamefont{Girvan}},
  \bibinfo{journal}{Statistical Mechanics of Complex Networks} pp.
  \bibinfo{pages}{66--87} (\bibinfo{year}{2003}).

\bibitem[{\citenamefont{Newman}(2004)}]{newman-fast}
\bibinfo{author}{\bibfnamefont{M.~E.} \bibnamefont{Newman}},
  \bibinfo{journal}{Physical Review E (Statistical, Nonlinear, and Soft Matter
  Physics)} \textbf{\bibinfo{volume}{69}}, \bibinfo{pages}{066133}
  (\bibinfo{year}{2004}).

\end{thebibliography}
